\documentclass[pra,twocolumn,showpacs]{revtex4}
\usepackage{amsmath,amssymb,mathrsfs}
\usepackage{psfrag}
\usepackage{graphicx}
\usepackage{graphics}
\usepackage{epsfig}
\usepackage{bm}
\usepackage{color}
\usepackage{verbatim,color,ulem}

\newcommand{\beq}{\begin{equation}}
\newcommand{\eeq}{\end{equation}}
\newcommand{\beqa}{\begin{eqnarray}}
\newcommand{\eeqa}{\end{eqnarray}}

\begin{document}

\title{{Finite-Range Corrections to the Thermodynamics \\ of 
	the One-Dimensional Bose Gas}}

\author{A. Cappellaro$^{1}$ and L. Salasnich$^{1,2}$}
\affiliation{$^{1}$Dipartimento di Fisica e Astronomia ``Galileo Galilei'' 
and CNISM, Universit\`a di Padova, via Marzolo 8, 35131 Padova, Italy \\
$^{2}$CNR-INO, via Nello Carrara, 1 - 50019 Sesto Fiorentino, Italy}

\begin{abstract}
The Lieb-Liniger equation of state accurately describes the zero-temperature 
universal properties of a dilute one-dimensional Bose gas in terms of 
the s-wave scattering length. For weakly-interacting bosons 
we derive non-universal corrections to this 
equation of state taking into account finite-range effects of the 
inter-atomic potential. Within the finite-temperature formalism of 
functional integration we find a beyond-mean-field equation of state 
which depends on scattering length and effective range of the 
interaction potential. Our analytical results, which are obtained 
performing dimensional regularization of divergent zero-point 
quantum fluctuations, show that for the one-dimensional Bose gas 
thermodynamic quantities like pressure and sound velocity 
{{are modified by changing the ratio between the 
effective range and the scattering length.}}
\end{abstract}

\pacs{03.70.+k, 05.70.Ce, 67.85.-d} 

\date{\today}

\flushbottom
\maketitle

\section{Introduction} 

For almost eighty years one-dimensional (1D) quantum 
systems have been subject of intense fascination. From the seminal work 
of Bethe in 1931 on the Heisenberg model \cite{bethe1931}, 
1D physics was then explored in great detail, 
providing exact and approximate solutions to a wide variety of 
systems \cite{giamarchi2003}. These 1D physical systems were considered 
toy models but recent advances in experimental setups, 
like Josephson nano-junctions \cite{chow1998,haviland2000} and 
magneto-optical trapping of cold atoms \cite{paredes2004,kinoshita2004,
haller2009,haller2010}, have achieved the realization of 1D quantum fluids. 
In the case of the 1D atomic Bose gas, Lieb-Lininger (LL) equation 
provides a reliable description of the zero temperature thermodynamic 
quantities \cite{lieb1963-1,lieb1963-2}, while numerical results at finite 
temperature can be extracted from the Yang-Yang theory 
\cite{yang1969,yang1970}, which reduces to LL one at zero temperature. 
The resulting thermodynamics is universal, since the only dependence from 
the inter-atomic potential is via the s-wave scattering length.
However, deviations from universality due to the finite-range of interaction 
potential have been shown to be quite important for a better understanding 
of the 3D Bose gas \cite{salasnich1998,braaten2001,roth2001,gao2003,
boris2003,andersen2004,pethick2007,zinner2009,ketterle2014,
sgarlata2015,cappellaro2017}. 
{{Finite-range effects naturally arise by modelling the two-body
interactions beyond the unphysical zero-range approximation.
As recently shown for the 2D Bose gas \cite{salasnich2017},
since finite-range corrections affect the behavior of quantum
and thermal fluctuations, they have to be taken into account also 
in lower dimensional system, where fluctuations are strongly enhanced
(Mermin-Wagner-Hohenberg theorem \cite{mermin1966,hohenberg1967})}.}
{{Improvements in cold-atom experimental techniques
are providing a precision benchmark which makes possible 
to explore in great details
beyond-mean-field effects in low-dimensional systems 
\cite{paredes2004,kinoshita2004,jochim2015-bkt,jochim2015}.}}
{{It was recently shown, both theoretically and experimentally,
that quantum fluctuations provides a stabilization mechanism against 
collapse, as predicted by the mean-field theory, both in 
dipolar condensate and binary Bose mixture
\cite{petrov2015,petrov2016,pfau2016,ferlaino2017,tarruell2017}. 
Moreover, by tuning 
the interaction via Feshbach resonances one can reach 
regimes where the effective range 
of interaction is comparable the scattering length: in these situations, a
universal GPE-based description is no more reliable, and one has to consider
finite-range effects in the thermodynamical and dynamical 
description of the system.
}}

In this paper we investigate the role of non-universal corrections 
to the thermodynamics of the 1D Bose gas both at zero and at finite 
temperature. Within the framework of a local effective action
\cite{braaten2001, roth2001,cappellaro2017,salasnich2017,braaten1999},
we derive a novel equation of state, where the contributions
of quantum and thermal fluctuations crucially depend also on the 
effective range of the interatomic potential. 
These corrections are computed at Gaussian (one-loop) level, 
where non-physical divergences are removed by using dimensional 
regularization. We show that the Gaussian theory  
reproduces extremely well the LL equation of state 
for weak and intermediate couplings. 
Finite-range effects are analyzed by considering 
measurable quantities such as the grand potential, the pressure, and the 
sound velocity. We find that, at fixed temperature and scattering length, 
these physical quantities can be widely tuned by varying the effective range. 

\section{Effective field theory for 1D Bose gases} 
  
The Euclidean Lagrangian density of identical bosonic 
particles of mass $m$ and chemical potential $\mu$ in a 1D configuration 
is given by \cite{altland2006}
\beq
\begin{aligned}
\mathcal{L} & = \psi^*(x,\tau) \bigg[\hbar\partial_{\tau} - \frac{\hbar^2}{2m}
\frac{\partial^2}{\partial x^2} -\mu \bigg]\psi(x,\tau) \\
& \quad \quad +\frac{1}{2}\int dx' |\psi(x',\tau)|^2 V(|x-x'|)
|\psi(x,\tau)|^2 \; ,
\end{aligned}
\label{lagrangian density}
\eeq
where bosons are described by the complex field $\psi(x,\tau)$ and 
$V(|x-x'|)$ is the two-body interaction potential between atoms.

For dilute and ultracold atomic gases the inter-atomic potential is usually 
approximated by a zero-range potential. 
In order to analyze the role played by the finite range of the inter-atomic
potential, we go beyond the zero-range approximation by considering
the following low-momentum expansion
\beq
\tilde{V}(q) = g_0 + g_2 q^2 + \mathcal{O}(q^4) 
\label{finite range pseudopotential}
\end{equation}
of the Fourier transform $\tilde{V}(q)$ of the interaction potential $V(|x|)$.
It is relevant to connect the parameters $g_0$ and $g_2$ with measurable 
quantities like the 1D scattering length $a_{s}$ and the 
1D effective range $r_e$ of the potential $V(|x|)$. In order to establish 
this connection, we first recall that, from 1D scattering theory, 
the scattering amplitude for an even scattering wavefunction
reads, in terms of the phase shift \cite{adhikari2000}
\beq
f_0(q) = q e^{i\delta_{0}(q)} \sin\delta_{0}(q)  \;.
\label{scattering ampitude}
\eeq
The scattering length and the effective of the inter-atomic potential
are defined via a low-momentum power expansion of the phase shift, namely
 \cite{adhikari2000}
\beq
q\tan\delta_{0}(q) = \frac{1}{a_{s}} + \frac{1}{2}r_e q^2 + \mathcal{O}(q^4) \;.
\label{effective range expansion}
\end{equation}
By supposing that the most relevant scattering processes in the system 
are the ones described  by $f_0(q)$, then \cite{stoof2009}, we then 
write the T-matrix as
\beq
 T_0(q) =  -\frac{2\hbar^2}{m}f_0(q)\;.
\label{scattering amplitude via t-matrix}
\eeq
At low momenta, the T-matrix is determined analytically by
solving the equation \cite{stoof2009,braaten2001}
\beq
T_0(q) = \bigg[\frac{1}{\tilde{V}(q)} - \frac{m}{2\pi \hbar^2}\int  
\frac{dp}{p^2 - q^2 + i\kappa}\bigg]^{-1}
\label{T-matrix eq.1}
\end{equation}
By taking $\tilde{V}(q)$ as in Eq. \eqref{finite range pseudopotential}, 
Eq. \eqref{T-matrix eq.1} becomes
\beq
T_0(q) = \bigg[\frac{1}{g_0} - \frac{g_2 q^2}{g_0^2} + \mathcal{O}(q^3) 
+ \frac{im}{2\hbar^2 q}\bigg]^{-1} \;.
\label{T-matrix with finite range}
\end{equation}
By inserting  Eq. \eqref{scattering ampitude} in 
Eq. \eqref{scattering amplitude via t-matrix}, with 
Eq. \eqref{effective range expansion}
and the identity $e^{i\delta}\sin \delta = 1/(\cot\delta -i)$,
an expansion up to $q^2$ also leads us to 
\beq
T_0(q) =-\frac{2\hbar^2}{m}\bigg[a_{s} -\frac{1}{2}
r_e a_{s}^2 q^2 + \mathcal{O}(q^3) -\frac{i}{q}\bigg]^{-1}\;.
\label{T-matrix with finite range v2}
\end{equation}
By matching Eq. \eqref{T-matrix with finite range} and 
Eq. \eqref{T-matrix with finite range v2}, we finally obtain  
\beq
g_0 = -\frac{2\hbar^2}{m a_{s}} \; \qquad \; g_2 = -\frac{\hbar^2}{m}r_e \;,
\label{expansion parameters}
\end{equation}
recovering the familiar result for the 1D coupling constant $g_0$ 
\cite{olshanii1998,cazalilla2011} and a remarkably simple 
formula relating $g_2$ to $r_e$. A similar procedure 
to get a relationship between $g_2$ and $r_e$ has been used 
for the 3D Bose gas \cite{braaten2001,cappellaro2017}, while in 2D 
the connection between $g_2$ and $r_e$ is much more complicated 
due to a logarithmic dependence on momentum in the T-matrix.

\section{Thermodynamic properties} 

All the relevant thermodynamic quantities 
can be derived from the grand canonical ensemble partition function
$\mathcal{Z}$ or, equivalently, from the grand potential $\Omega$, 
which are defined as
\beq
\mathcal{Z}= e^{-\beta \Omega} = \int \mathcal{D}[\psi,\psi^*]
\exp\bigg(-\frac{1}{\hbar}
S[\psi,\psi^*]\bigg) \;
\label{partition function}
\end{equation}
adopting a functional integration formalism.
In Eq. \eqref{partition function} the Euclidean action is given by
\beq
S[\psi,\psi^*] = \int_L dx\int_0^{\beta\hbar}d\tau \,\mathcal{L}
\big[\psi,\psi^*\big]\;,
\label{euclidean action}
\end{equation}
where $\mathcal{L}[\psi,\psi^*]$ is defined in Eq. 
\eqref{lagrangian density}, $L$ is the system size and 
$\beta \equiv 1/(k_B T)$, with $k_B$ the Boltzmann constant 
and $T$ the temperature.

The mean-field plus Gaussian approximation is obtained by splitting the
field $\psi(x,\tau)$ as
\beq
\psi(x,\tau) = v + \eta(x,\tau)
\label{splitting}
\eeq
and expanding the action $S[\psi,\psi^*]$ around the constant field $v$ 
up to the quadratic terms in the fluctuations field $\eta$ and $\eta^*$. 
{{Within the framework of second quantization, this procedure 
is closely related to the Bogoliubov approximation,
where the quartic hamiltonian is disentangled in quadratic terms
\cite{fetterbook}.}}
The grand potential is then composed by three different contribution
\beq
\Omega (\mu,v,T) = \Omega_{0}(\mu,v) + 
\Omega_g^{(0)}(\mu,v) + \Omega^{(T)}_g(\mu,v,T)\;.
\label{grand potential terms}
\end{equation}
Assuming a real $v$, $\Omega_{0}(\mu,v)$ 
is given by the terms in the action independent 
from $\eta$ and $\eta^*$, namely \cite{toigo2016}
\beq
\Omega_0(\mu,v) = L\ \Big(-\mu v^2
+ \frac{1}{2}g_0 v^4\Big)\;
\label{Omega0 with mu + psi}
\eeq
The contribution of fluctuations  has double nature: first, quantum 
fluctuations arising a zero temperature, i.e. the zero-point energy 
of collective excitations
\beq
\Omega_g^{(0)} = \frac{1}{2}\sum_q E_q(\mu,v)
\label{quantum fluctuations}
\end{equation}
and then the thermal ones
\beq
\Omega_g^{(T)} = \beta^{-1}\sum_q \log\big(1 - e^{-\beta E_q(\mu,v)} \big) \;.
\label{thermal fluctuations}
\end{equation}
In Eq. \eqref{quantum fluctuations} and Eq. \eqref{thermal fluctuations}, 
$E_q(\mu,v)$ is the spectrum of bosonic excitations:
\beq
E_q(\mu,v) = \bigg[ \bigg(\frac{\hbar^2 q^2}{2m} -
\mu+v\,^2[g_0 + \tilde{V}(q)] \bigg)^2 
 -v\,^4\tilde{V}(q)^2  \bigg]^{1/2} \;.
 \label{collective excitations}
\end{equation}
One can eliminate the dependence on the parameter $v$ by imposing the 
saddle-point condition on $\Omega_{0}(\mu,v)$, which reads 
\beq
v = \sqrt{\frac{\mu}{g_0}}\;,
\label{saddle point}
\end{equation}
and make the spectrum in Eq. \eqref{collective excitations} gapless
\beq
E_q(\mu) = \sqrt{\frac{\hbar^2 q^2}{2m} 
\bigg[ (1+\chi \mu)\frac{\hbar^2 q^2}{2m} + 2\mu\bigg]}\;,
\label{collective excitations gapless}
\end{equation}
where finite range effects are encoded in the parameter
\beq
\chi = \frac{4m}{\hbar^2}\frac{g_2}{g_0} \;.
\label{parameter chi}
\end{equation}
The grand potential $\Omega(\mu,T)$ is easily related to the pressure
through the formula $P= - \Omega / L$ so,
by replacing Eq. \eqref{saddle point} in Eq. \eqref{Omega0 with mu + psi}, 
one gets the familiar mean-field result
\beq
P_{0}(\mu) = \frac{\mu^2}{2g_0} \; , 
\label{Omega0 solo mu}
\eeq
which reproduces the LL solution in the weak-coupling regime. 
Notice that this mean-field contribution depends only on the 
zero-range strength $g_0$. 

\subsection{Quantum fluctuations} 

In the continuum limit the contribution arising 
from quantum fluctuations is given by 
\beq
P^{(0)}_g(\mu) = - \frac{1}{2}\int_{-\infty}^{+\infty} 
\frac{dq}{2\pi} E_q (\mu)\;.
\label{pressure definition continuum}
\end{equation}
This integral is ultraviolet divergent. In order to regularize it 
we use dimensional regularization \cite{hooft1972,phillips1997}: 
the integral of $E_q(\mu)$ is computed over a $D$-dimensional 
momentum space, finding 
\beq
P^{(0)}_g = -\frac{S_D(2\tilde{\mu})^{D/2 +1}}{4(2\pi)^D}
\Big(\frac{2m}{\hbar^2}\Big)^{D/2} B\Big(\frac{D+1}{2},-\frac{D+2}{2} \Big) \; ,
\label{regolarizzazione dimensionale}
\end{equation}
where $\tilde{\mu} = \mu/(1+\chi \mu)$, 
$S_D=2\pi^{D/2}/\Gamma(D/2)$ with $\Gamma(x)$ the Euler Gamma function, 
and $B(x,y)$ the Euler beta function. The dimension $D$ is taken complex 
such that $B(x,y)$ can be analytically continued and expressed 
in terms of Gamma functions: $B(x,y)=\Gamma(x)\Gamma(y)/\Gamma(x+y)$. 
Finally, setting $D=1$ we obtain 
\beq
P_g^{(0)}(\mu) = \frac{2}{3\pi}\sqrt{\frac{m}{\hbar^2}} \ 
\frac{\mu^{3/2}}{1+\chi \mu} \; .
\label{zero temperature granpot}
\eeq
This beyond-mean-field Gaussian contribution 
depends explicitly only on the finite-range parameter $\chi$, 
given by Eq. \eqref{parameter chi}. 

The zero-temperature number density $n$ of the system is 
immediately derived from 
\beq
n = {\partial \over \partial \mu} \left( P_0(\mu) + P_g^{(0)}(\mu) \right) \; 
\label{zero temperature density} 
\eeq
and the corresponding speed of sound reads 
\beq
c_s = \sqrt{\frac{n}{m} \frac{\partial \mu}{\partial n}}\; . 
\label{sound velocity definition}
\eeq

\begin{figure}[ht]
\centering
\includegraphics[width=8.5cm,clip=]{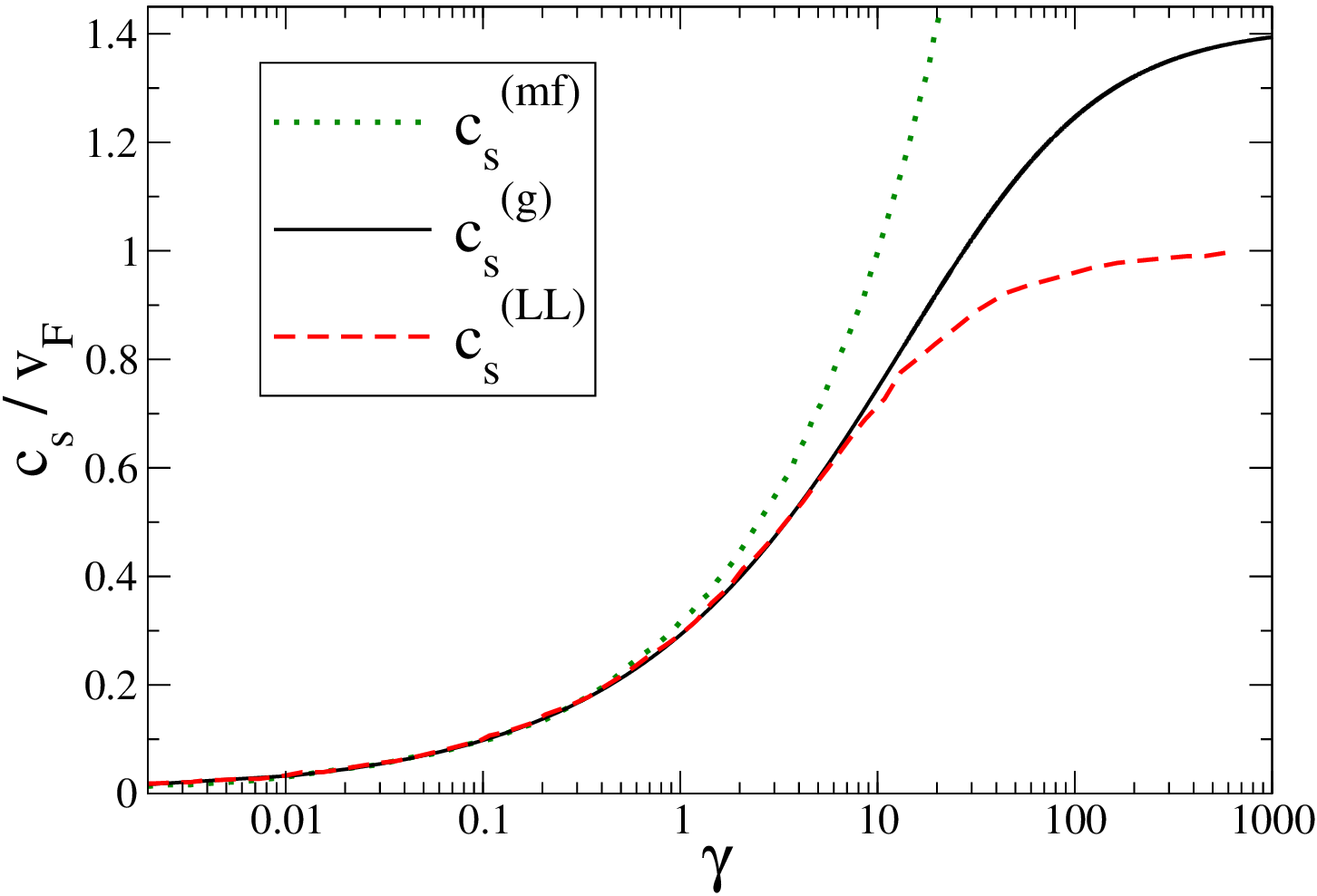}
\includegraphics[width=8.5cm,clip=]{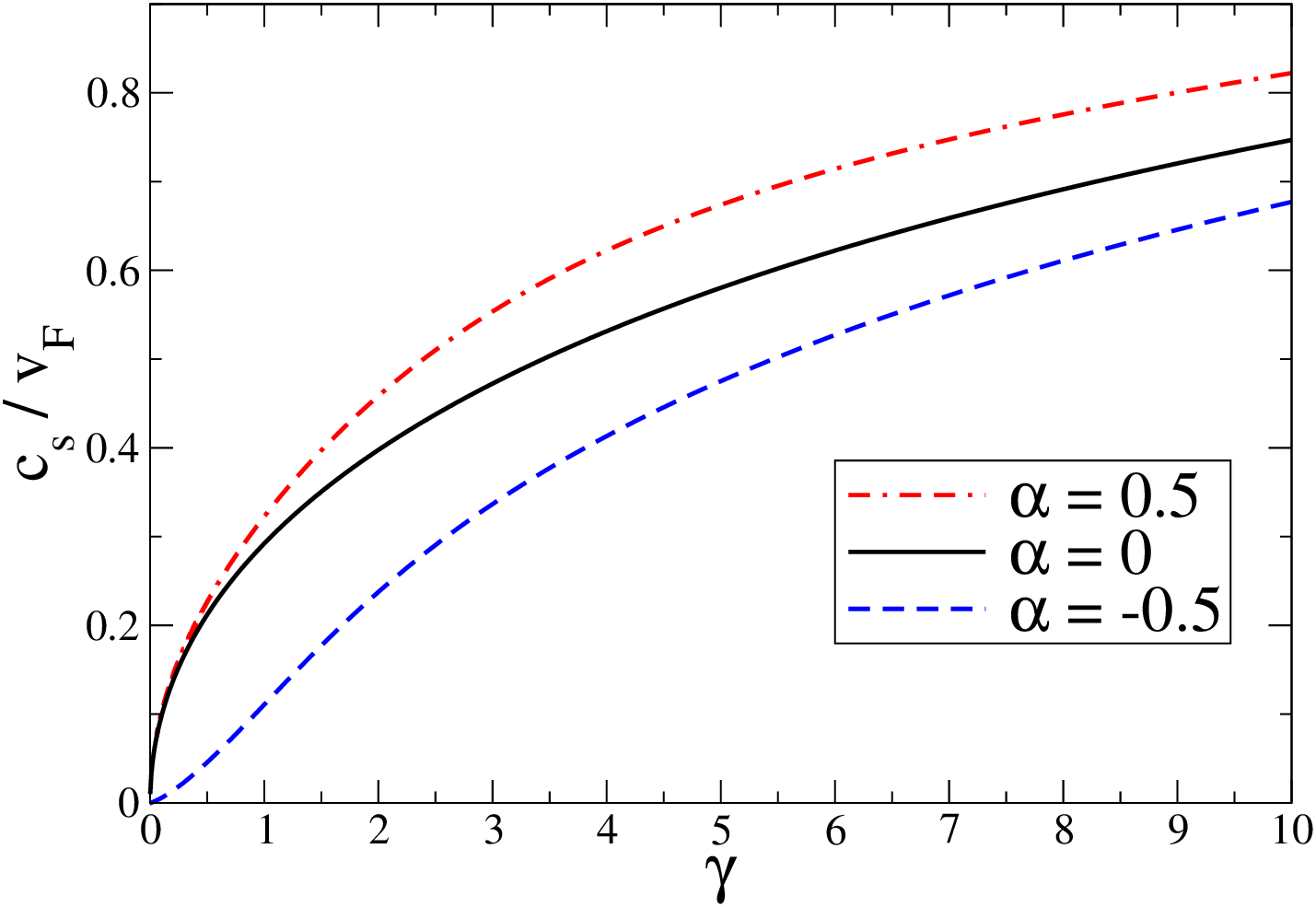}
\caption{Upper panel: Zero-temperature sound velocity in units of 
the Fermi velocity $v_F = \hbar\pi n / m $ as a function of 
the adimensional zero-range interaction parameter 
$\gamma = (mg_0)/(\hbar^2 n)$ \cite{kaminaka2011,stringari2017}. 
We compare our Gaussian sound velocity $c_s^{(g)}$ (solid line), 
with the exact solution of LL equation \cite{stringari2017} 
(dashed line), and the mean-field result $c_s^{(mf)}$ (dotted line). 
Lower panel: Gaussian sound velocity for different values of the 
adimensional parameter $\alpha = \hbar^2\chi/(m a_s^2)=4 r_e/a_{s}$, 
which encodes the role played by the finite range 
of the inter-atomic potential.} 
\label{fig1}
\end{figure}

In the upper panel of Fig. \ref{fig1} we report 
the comparison between our Gaussian result of the sound velocity 
(solid line) and the solution of LL theory (dashed line) for the universal case 
($\chi =0$ in Eq. (\ref{zero temperature granpot})). 
We characterize the zero-range interaction strength by the parameter 
$\gamma = (mg_0)/(\hbar^2 n)$ \cite{cazalilla2011,stringari2017}. 
The figure shows that our Gaussian theory improves the mean-field result, 
$c_s^{(mf)}(\gamma) = \hbar n\sqrt{\gamma}/m$, 
reliable only in the weak-coupling limit (dotted line). 
Quite remarkably, the range of applicability of our Gaussian theory 
is up to $\gamma \simeq 10$. Only the very strong coupling 
(Tonks-Girardeau) regime of impenetrable bosons 
is not captured by the Gaussian theory. 

In the lower panel of Fig. \ref{fig1} we plot the 
behavior of our Gaussian sound velocity for different values of the 
finite-range adimensional parameter $\alpha = (\hbar^2 \chi)/(m a_{s}^2) 
=4 r_e/a_{s}$. 
The figure clearly shows that the inclusion of finite range effects 
deeply affects the zero-temperature sound velocity: 
at fixed zero-range strength the sound velocity becomes larger 
with a positive effective range, while the opposite happens 
for a negative effective range. 
As previously stated, enhanced quantum fluctuations play
a crucial role in low-dimensional atomic quantum gases 
\cite{mermin1966,hohenberg1967}, 
since they cannot be controlled by lowering the temperature. Thus, 
the usual mean-field scheme has to be modified in such a way to include
Gaussian fluctuations in the thermodynamic and dynamical description provided
respectively by the grand potential $\Omega(\mu)$ and the Gross-Pitaevskii 
equation (GPE). 

It has been shown that these corrections
deeply affects the stability of 2D and 1D Bose-Bose mixtures, enabling 
a transition from a homogeneous ground state 
to liquid-like self-bound states \cite{petrov2015,petrov2016}. 		
Moreover, while a universal GPE-based approach is 
reliable only when $r_e \ll a_s \ll d$,
with $d$ being the average distance between atoms, it has been shown 
\cite{boris2003,pethick2007,ketterle2014}
that one can greatly enhance finite-range effects by means of 
Fano-Feshbach resonances. In regimes where $r_s \approx a_s$, deviation 
from universalities are relevant, and every dynamical picture of the system has
to take into account not only quantum fluctuations, but also finite-range 
corrections modelled around Eq. \eqref{zero temperature granpot}. 
In the present experiments one can tune the scattering length $a_s$ 
and in turn the adimensional parameter $\alpha= 4 r_e/a_s$ 
of our theory, because the effective range $r_e$ remains 
practically fixed and quite small ($\simeq 10^{-10}$ m). 
The nontrivial behavior of the sound velocity $c_s$ shown 
in the lower panel of Fig. \ref{fig1} can be observed experimentally, 
at fixed $a_s$ and $r_e$, by varying the number density $n$ and 
consequently the adimensional interaction parameter $\gamma$.
{{
Our approach can be seen as complementary to the one in \cite{choi2015}, 
where collective excitations of the 1D Bose gas are studied 
within a hydrodynamic framework based on a GP-like equation. 
Differently from the usual mean-field approach,
they still consider a zero-range two-body interaction, but  
the chemical potential is modelled around the LL exact solution.}}

\subsection{Thermal fluctuations} 

The contribution of thermal fluctuations 
is obtained from Eq. \eqref{thermal fluctuations}. 
In the continuum limit 
\beq
P^{(T)}_g(\mu,T) = \frac{1}{\pi}\int_0^{+\infty}\; dq\,
q\frac{\text{d}E_q}{\text{d}q} \bigg(\frac{1}{e^{\beta E_q} - 1}\bigg) \; .
\label{finite temp. eq. 1}
\eeq

\begin{figure}[ht]
\centering
\includegraphics[width=8.5cm,clip=]{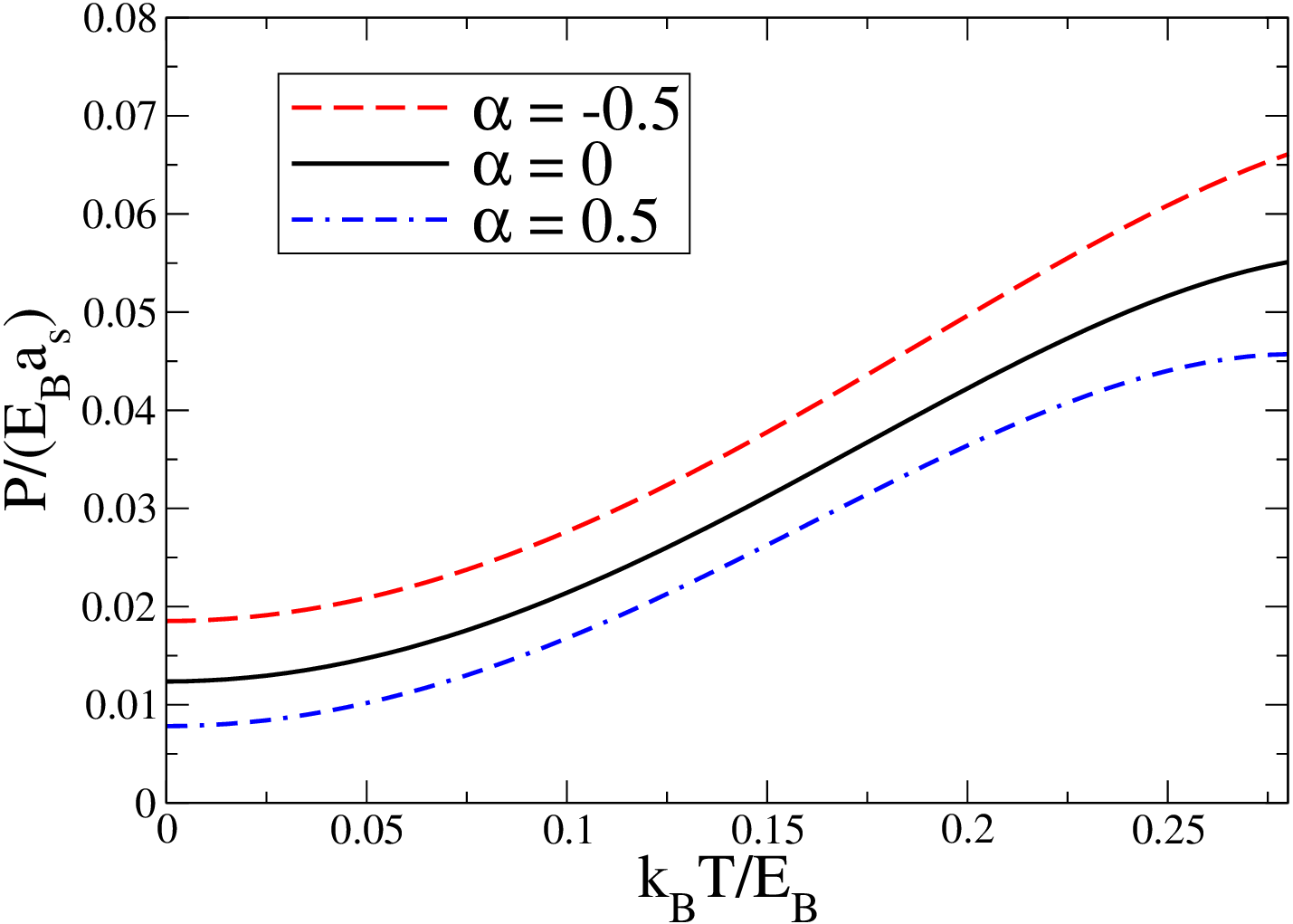}
\caption{We plot the pressure $P(\mu,T)$ in Eq. 
\eqref{bmf pressure finite temperature},
with $k_B T$ in units, and $\mu$ in units of $E_B = \hbar^2/(ma_{s}^2)$, 
at the fixed value $\mu/E_B = 0.3$. We consider three different values 
of the adimensional parameter $\alpha = 4r_e/a_{s}$. The case 
$\alpha = 0$ corresponds to the zero-range approximation of the 
inter-atomic potential.} 
\label{fig2}
\end{figure}

By changing the integration variable to $x=\beta E_q(\mu)$, one gets
\beq
P^{(T)}_g(\mu,T)  = \frac{1}{\pi\beta} \int_0^{+\infty} dx \;
q(x) \frac{1}{e^x - 1}\;
\label{finite temp eq. 2}
\eeq
where
\beq
q(x) = \sqrt{\frac{2m\mu}{\hbar^2(1+\chi \mu)}}\sqrt{-1+ \sqrt{1 + 
\frac{(1+\chi \mu) x^2}{\mu^2}(k_BT)^2}}\;.
\label{definition of q(x)}
\eeq
After having inserted Eq. \eqref{definition of q(x)} in 
Eq. \eqref{finite temp eq. 2}, an expansion at low temperatures finally 
leads us to the following results 
\beq
P^{(T)}_g(\mu,T) = \frac{\pi}{6}\sqrt{\frac{m}{\hbar^2 \mu}} (k_B T)^2 
\bigg[ 1 - \frac{\pi^2}{20}\frac{1+\chi \mu}{\mu^2}(k_B T)^2\bigg] \; . 
\label{finite temperature final results} 
\eeq
So, by recalling Eq. \eqref{Omega0 solo mu} and 
Eq. \eqref{zero temperature granpot}, the gran canonical ensemble 
pressure is given by 
\beq
P(\mu,T) = P_0(\mu) + P_g^{(0)}(\mu) + P_g^{(T)}(\mu, T)\;.
\label{bmf pressure finite temperature}
\eeq 
In Fig. \ref{fig2} we report the pressure $P$ as a 
function of temperature $T$ at fixed chemical potential $\mu$
for three values of the finite-range 
adimensional parameter $\alpha = 4r_e/a_{s}$, 
obtained with the low-temperature analytical formula 
(\ref{finite temperature final results}). 
At finite temperature non-universal effects slightly 
increase as ratio $(k_BT/\mu)^4$ grows. This is explained 
by recalling that the details of the inter-atomic potential 
become more relevant when atoms scatter at higher energy.

\section{Conclusions} 

We have derived a non-universal equation of state for the 1D Bose gas  
providing a truthful connection between the scattering parameters 
and the coupling constants of the interaction potential. 
The non-universal 1D equation of state is quite different 
with respect to the 3D \cite{braaten2001} and 2D \cite{salasnich2017} ones,   
and its quantum fluctuations crucially depend on the scattering 
length and the effective range of the interaction potential. 
Our finite-range analytical results, 
Eq. \eqref{zero temperature granpot} and Eq. 
\eqref{finite temperature final results}, represent a nontrivial 
generalization of the universal 1D equation of state. 
In the case of contact interaction 
we have shown that our beyond-mean-field theory reproduces 
extremely well the Lieb-Liniger one apart the very strong coupling 
regime. Sizable non-universal effects can be reached experimentally 
with confined quasi-1D ultracold atoms through Feshbach-resonance techniques 
by decreasing the 1D scattering length $a_s$. 
Indeed, finite-range effects become relevant by increasing 
the adimensional ratio $r_e/a_s$, where $r_e$ 
is the 1D effective range of the inter-atomic potential. 
Our beyond-mean-field description of the 1D Bose gas 
can help the understanding of a wide range of problems. 
For instance, the role of quantum fluctuations, thermal fluctuations 
and finite-range corrections in the stability and structural properties  
of topological defects (dark solitons) and 
localized self-bound states (i.e. bright solitons).

{\bf Acknowledgments}. 
The authors acknowledge for partial support the 2016 BIRD project 
"Superfluid properties of Fermi gases in optical potentials" of the 
University of Padova. The authors thank Giacomo Bighin and 
Flavio Toigo for enlightening discussions.


\begin{thebibliography}{99}

\bibitem{bethe1931} H. Bethe, Z. Physik \textbf{71}, 205 (1931).

\bibitem{giamarchi2003} T. Giamarchi, \textit{Quantum Physics in One
Dimension}, (Clarendon Press, Oxford, 2003).

\bibitem{chow1998} E. Chow, P.  Delsing and D. B. Haviland, 
Phys. Rev. Lett. \textbf{81}, 204 (1998).

\bibitem{haviland2000} D. B. Haviland, K. Andersson, and P. Agren, J. Low Temp.
Phys. \textbf{118}, 733 (2000).

\bibitem{paredes2004} B. Paredes, A.Widera, V. Murg, O.Mandel, 
S. Folling, I. Cirac, G. V. Shlyapnikov, T.W. Hansch, and I. Bloch, 
Nature (London) \textbf{429}, 277 (2004).

\bibitem{kinoshita2004} T. Kinoshita, T. Wenger, and D. S. Weiss, 
Science \textbf{305}, 1125 (2004).

\bibitem{haller2009}E. Haller, M. Gustavsson, M. J. Mark, J. G. Danzl, R. Hart,
G. Pupillo, and H.-C. Nägerl, Science \textbf{325}, 1224 (2009).

\bibitem{haller2010} E. Haller, R. Hart, M. J. Mark, J. G. Danzl, 
L. Reichsoller, M. Gustavsson, M. Dalmonte, G. Pupillo, and H.-C. Nagerl, 
Nature (London) \textbf{466}, 597 (2010).

\bibitem{lieb1963-1} E. H. Lieb and W. Lininger, 
Phys. Rev. \textbf{130}, 1605 (1963).

\bibitem{lieb1963-2} E.H. Lieb and W. Lininger, Phys. Rev. 
\textbf{130}, 1616 (1963).

\bibitem{yang1969} C. N. Yang and C. P. Yang, J. Math. Phys. 
\textbf{10}, 1115 (1969).

\bibitem{yang1970} C. N. Yang and C. P. Yang, Phys. Rev. A 
\textbf{2}, 154 (1970).

\bibitem{salasnich1998} A. Parola, L. Salasnich, and L. Reatto, 
Phys. Rev. A {\bf 57}, R3180 (1998). 

\bibitem{braaten2001} E. Braaten, H.-W. Hammer, and S. Hermans, 
Phys. Rev. A \textbf{63}, 063609 (2001).

\bibitem{roth2001} R. Roth and H. Feldmeier, Phys. Rev. A 
\textbf{64}, 043603 (2001).

\bibitem{gao2003} H. Fu, Y. Wang, and B. Gao, Phys. Rev. A 
\textbf{67}, 053612 (2003). 

\bibitem{boris2003} J.J. Garcia-Ripoll, V.V. Konotop, B. A. Malomed, 
and V.M. Perez-Garcia, Mathematics and Computers in Simulation {\bf 62}, 
21 (2003).

\bibitem{andersen2004} J. O. Andersen, Rev. Mod. Phys. \textbf{76}, 599 (2004).

\bibitem{pethick2007} A. Collin, P. Massignan, and C.J. Pethick, 
Phys. Rev. A {\bf 75}, 013615 (2007).

\bibitem{zinner2009} N.T. Zinner and M. Thogersen, 
Phys. Rev. A {\bf 80}, 023607 (2009); M. Thogersen, 
N.T. Zinner, and A.S. Jensen, Phys. Rev. A {\bf 80}, 043625 (2009). 

\bibitem{ketterle2014} H. Veksler, S. Fishman, and W. Ketterle, 
Phys. Rev. A {\bf 90}, 023620 (2014). 

\bibitem{sgarlata2015} F. Sgarlata, G. Mazzarella, and L. Salasnich, 
J. Phys. B: At. Mol. Opt. Phys. \textbf{48}, 115301 (2015).

\bibitem{cappellaro2017} A. Cappellaro and L. Salasnich, Phys. Rev. A 
\textbf{95}, 033627 (2017).

\bibitem{salasnich2017} L. Salasnich, Phys. Rev. Lett. \textbf{118}, 
130402 (2017).

\bibitem{mermin1966} N. D. Mermin and H. Wagner, Phys. Rev. Lett. 
\textbf{17}, 1133 (1966).

\bibitem{hohenberg1967} P. C. Hohenberg, Phys. Rev. \textbf{158}, 383 (1967).

{{
\bibitem{jochim2015-bkt} P. A. Murthy, I. Boettcher, L. Bayha, 
M. Holzmann, D. Kedar, M. Neidig, M. G. Ries, A. N. Wenz, G. Zürn
and S. Jochim, Phys. Rev. Lett. \textbf{115}, 010401 (2015).
		
\bibitem{jochim2015} S. Murmann, F. Deuretzbacher, G. Zurn,
J. Bjerlin, S. M. Reimann, L. Santos, T. Lompe, and S. Jochim,
Phys. Rev. Lett. \textbf{115}, 215301 (2015).}

\bibitem{petrov2015} D.S. Petrov, Phys. Rev. Lett. \textbf{115}, 155302 (2015).

\bibitem{petrov2016} D.S. Petrov, Phys. Rev. Lett. \textbf{117}, 100401 (2016).

\bibitem{pfau2016} M. Schmitt, M. Wenzel, F. Böttcher, 
I. Ferrier-Barbut and Tilman Pfau, Nature \textbf{539}, 259 (2016).

\bibitem{ferlaino2017} L. Chomaz, S. Baier, D. Petter, M. J. Mark, 
F. Wächtler, L. Santos, F. Ferlaino, Phys. Rev. X \textbf{6}, 041039 (2017).

\bibitem{tarruell2017} C. R. Cabrera, L. Tanzi, J. Sanz, B. Naylor, 
P. Thomas, P. Cheiney, L. Tarruell, arXiv:1708.07806 (2017).
}

\bibitem{braaten1999} E. Braaten and A. Nieto, Eur. Phys. J. B \textbf{11}, 
143 (1999).

\bibitem{altland2006} A. Altland and B. Simons, \textit{Condensed Matter 
Field Theory} (Cambridge University Press, Cambridge, England, 2006).

\bibitem{adhikari2000} V. E. Barlette, M. M. Leite, and S. Adhikari, 
Eur. J. Phys. \textbf{21},  435-440 (2000).

\bibitem{stoof2009} H. T. C. Stoof, D. B. M. Dickerscheid, and K. Gubbels,
\textit{Ultracold Quantum Fields}. 
(Springer, Dordrecht, 2009).

\bibitem{olshanii1998} M. Olshanii, Phys. Rev. Lett. \textbf{81}, 938 (1998). 

\bibitem{cazalilla2011} M. A. Cazalilla, R. Citro, T. Giamarchi, 
E. Orignac and M. Rigol, Rev. Mod. Phys. \textbf{83}, 1405 (2011).

{{
\bibitem{fetterbook} A. L. Fetter and J. D. Walecka, \textit{Quantum
Theory of Many-Particle Systems} (McGraw-Hill, Boston, 1971). }}

\bibitem{hooft1972} G. 't Hooft and J. M. G. Veltman, 
Nucl. Phys. B \textbf{44}, 189 (1972).

{
{\bibitem{phillips1997} D. R. Phillips, S. R. Beane and T. D. Cohen, 
Annals Phys. \textbf{263}, 255-275 (1998).}}

\bibitem{toigo2016} L. Salasnich and F. Toigo, 
Phys. Rep. \textbf{640}, 1 (2016).

\bibitem{stringari2017} G. De Rosi, G. E. Astrakharchik, and S. Stringari, 
Phys. Rev. A \textbf{96}, 013613 (2017).

\bibitem{kaminaka2011} T. Kaminaka and M. Wadati, Phys. Lett. A 
\textbf{375}, 2460 (2011).

{{
\bibitem{choi2015} S. Choi, V. Dunjko, Z.D. Zhang, M. Olshanii,
Phys. Rev. Lett. \textbf{115}, 115302 (2015).
}}


\end{thebibliography}
\end{document}